\documentclass{aipproc}
\usepackage{epsfig}
\layoutstyle{6x9}
\begin{document}
\title{Non-perturbative Aspects of Schwinger-Dyson Equations}
\author{A. Bashir}{
address={Instituto de F{\'\i}sica y Matem\'aticas, 
         Universidad Michoacana de San Nicol\'as de Hidalgo,
         Apartado Postal 2-82, Morelia, Michoac\'an 58040, M\'exico.}
}
\begin{abstract} 

Schwinger-Dyson equations (SDEs) provide a natural staring point to
study non-perturbative phenomena such as dynamical chiral symmetry
breaking in gauge field theories. We briefly review this research in the
context of quenched quantum electrodynamics (QED) and discuss the
advances made in the gradual
improvement of the assumptions employed to solve these equations.
We argue that these attempts render the corresponding studies
more and more reliable and suitable for their future use in the more
realistic cases of unquenched QED, quantum chromodynamics (QCD) and
models alternative to the standard model of particle physics. 

\end{abstract}
\maketitle

\newcommand{\be}{\begin{eqnarray}}
\newcommand{\ee}{\end{eqnarray}}

\section*{Introduction}

The standard model of particle physics is highly successful in collating 
experimental
information on the basic forces. Yet, its key parameters, the masses
of the quarks and leptons, are theoretically undetermined. In the
simplest version of the model, these masses are specified by the
couplings of the Higgs boson, couplings that are in turn undetermined.
However, it could be that it is the dynamics of the fundamental gauge
theories themselves that generate the masses of all the matter fields.
To explore this possibility, the favorite starting point is to
consider quenched QED as the simplest example of a
gauge theory and study the behavior of the fermion propagator, using
the corresponding SDE. Apart from the fermion propagator itself, the
only unknown ingredient in this equation is the fermion-boson vertex.
As the SDE of the vertex is quite complicated, a common
practice is to start from a suitable construction for it. One should
ensure that every {\em ansatz} of a non-perturbative fermion-boson 
interaction must have the following characteristics~:

\begin{itemize}

\item

It should respect the Ward-Green-Takahashi identity (WGTI) which relates it
to the fermion propagator. Moreover, in the limit when the fermion momenta 
are identical, it should also obey the limiting Ward identity (WI).

\item

In the weak coupling regime, it should match onto its perturbative loop
expansion.

\item

 It should transform according to the Landau-Khalatnikov-Fradkin 
transformations (LKFT) under a variation of gauge. Moreover, it must
guarantee that when used in the SDE for the fermion propagator, the
resulting propagator also obeys its corresponding LKFT.

\item

It should not contain any kinematic singularities. 

\item

If we are studying 3+1-dimensional QED, it should ensure that the fermion
propagator is multiplicatively renormalizable.

\item

It should render the physical observables associated with the fermion
propagator gauge independent.

\end{itemize}

In addition to these factors, it is also important to solve the SDE for
the fermion propagator by employing a gauge invariant regulator. Since 
the earliest works on the dynamical breakdown of chiral symmetry
through the SDEs, \cite{Maskawa}, a lot of research has been
carried out in order that the above-mentioned goals could be achieved.
We review this work in the next sections after a brief overview of
the SDE for the fermion propagator.

\section*{SDE for the fermion propagator}

The SDE for the fermion propagator, $S_{F}(p)$,
in QED with a bare
coupling, $e$, is displayed in Fig.~(1), and is given by~:
\begin{center}
\begin{figure}[t]
\includegraphics{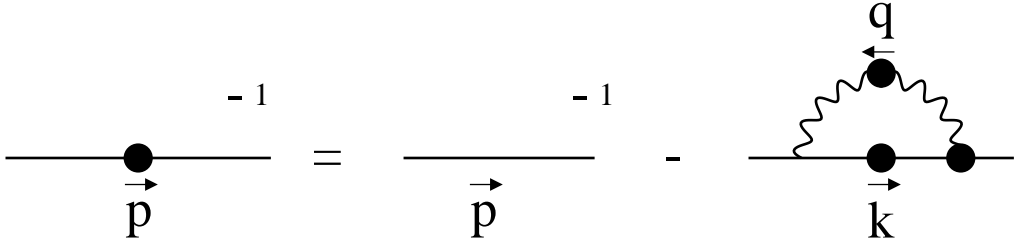}
\vspace{2mm}
\caption{Schwinger-Dyson equation for the fermion propagator.}
\label{SDEprop}
\end{figure}
\end{center}
\vspace{-5mm}
\begin{eqnarray}
     iS_{F}^{-1}(p)\,=\,iS_{F}^{0^{-1}}(p)\,-\,e^2\int \frac{d^4k}{(2\pi)^4}\,
    \gamma^{\mu}\,S_{F}(k)\, \Gamma^{\nu}(k,p)\,\Delta_{\mu\nu}
    (q) \;,   \label{SDE}
\end{eqnarray}
where  $q=k-p$ and $S_{F}(p)$ can be expressed in terms of two Lorentz scalar
functions, $F(p^2)$, the wavefunction renormalization, and
${\cal M}(p^2)$, the mass function, so that
\begin{eqnarray}
    S_{F}(p)&=& \frac{F(p^2)}{\not \hspace{-0.4mm} \! p- {\cal M}(p^2)} \; .
\end{eqnarray}
The bare propagator $ S_{F}^{0}(k)= 1/ (\not\hspace{-0.4mm}  
\! p - m_{0}) $, where $m_0$
is the constant (bare) mass. In quenched QED, the photon propagator is
unrenormalized and so is given by its bare form~:
\begin{eqnarray*}
 \Delta_{\mu\nu}(q) \equiv \Delta_{\mu\nu}^{0}(q) =\,\frac{1}{q^2}\ \left(
    g_{\mu\nu}+(\xi-1)\frac{q_{\mu}q_{\nu}}{q^2}\ \ \right) \; .
\end{eqnarray*}
$\Gamma^{\mu}(k,p)$ is the full fermion-boson vertex. Once it is known,
one can solve Eq.~(\ref{SDE}) to fully determine the fermion propagator
in terms of $F(p^2)$ and ${\cal M}(p^2)$. From these quantities, one can
extract physical observables such as the dynamically generated mass of
the fermion, the condensate $<\bar{\psi} \psi>$ and the critical
coupling above which the chiral symmetry is broken.

\section*{Ward-Green-Takahashi Identity}

The simplest {\em ansatz} for the full vertex is the bare vertex, i.e., 
$\Gamma^{\mu}(k,p)=\gamma^{\mu}$. Making use of it in Eq.~(\ref{SDE}),
we solve the latter for $F(p^2)$ and ${\cal M}(p^2)$ after setting 
$m_0=0$. We find a 
non-trivial solution for ${\cal M}(p^2)$ (different from the trivial solution 
${\cal M}(p^2)=0$ obtained in perturbation theory) at a value higher than
a critical value
of coupling $\alpha=\alpha_c$. In other words, above $\alpha_c$, fermions 
become massive
and below this value, they remain massless. As $\alpha_c$ corresponds to
a change of phase, we expect it to be a gauge independent quantity.
However, if one solves Eq.~(\ref{SDE}) for different values of the
gauge parameter $\xi$, one finds $\alpha_c$ highly gauge dependent. 
As a consequence of gauge covariance, Green
functions obey certain identities which relate one function to
the other. These relations have been named Ward-Green-Takahashi identities
(WGTI), \cite{WGT}. At the level of physical observables, gauge symmetry
reflects as the fact that they be independent of
the gauge parameter. Owing to the fact that the bare vertex does not
satisfy the WGTI $q_{\nu} \Gamma^{\nu}(k,p) = S_F^{-1}(k) -  S_F^{-1}(p)$
beyond the first order in perturbation theory, we cannot expect the
physical observables borne out of this approximation to be gauge
independent. In order to incorporate WGTI into the {\em ansatz} for the 
vertex, we follow Ball and Chiu, \cite{BC}, and write out the vertex as a sum of 
longitudinal and transverse components~:
\begin{eqnarray}
      \Gamma^{\mu}(k,p)=\Gamma^{\mu}_{L}(k,p)+\Gamma^{\mu}_{T}(k,p)\; .
\end{eqnarray}
By definition, the transverse part $\Gamma^{\mu}_{T}(k,p)$ satisfies 
$q_{\mu} \Gamma^{\mu}_{T}(k,p)=0$ and is undetermined the WGTI.
It also satisfies $\Gamma^{\mu}_{T}(p,p)=0$.
Ball and Chiu suggest the following longitudinal part in order to
satisfy WGTI in a manner free of kinematic singularities~:
\begin{eqnarray}
\Gamma^{\mu}_{L}(k,p)= a(k^2,p^2) \gamma^{\mu}
                       + \mbox{} b(k^2,p^2) (\not \! k + \not \! p)
                           (k+p)^{\mu}
                       - \mbox{} c(k^2,p^2) (k+p)^{\mu} \;,
\end{eqnarray}
 where,
\begin{eqnarray}\nonumber
  a(k^2,p^2)&=&\frac{1}{2}\ \left( \frac{1}{F(k^2)}\ +
  \frac{1}{F(p^2)}\  \right) \; , \\
  b(k^2,p^2)&=&\frac{1}{2}\ \left(
  \frac{1}{F(k^2)} - \frac{1}{F(p^2)}\ \right) \frac{1}{k^2 - p^2} \; ,  \\
\nonumber
  c(k^2,p^2)&=&  \left( \frac{{\cal M}(k^2)}{F(k^2)}\
  - \frac{{\cal M}(p^2)}{F(p^2)}\ \right) \frac{1}{k^2 - p^2} \;.
\end{eqnarray}
Though the choice of the longitudinal part of the vertex is not 
unique\footnote{Some other attempts to construct the longitudinal vertex 
can be found in references \cite{Atkinson,Haeri}.},
one of the advantages of the {\em ansatz} proposed by Ball and Chiu is that it
contains no kinematic singularities. Moreover, they also propose a
basis of eight independent tensors to write out the transverse part of
the vertex~:
\be
     \Gamma^{\mu}_T(k,p) &=& \sum_{i=1}^{8} \tau_i(k,p) \; T_i^{\mu}(k,p) \;.
\ee
They construct their basis such that the coefficients of
each of the basis vectors is independent of kinematic singularities
at the one loop level in the Feynman gauge \footnote{A complete one loop 
calculation of the transverse vertex in an arbitrary covariant gauge, 
\cite{Ayse}, slightly modifies this basis.}. It is a common
practice to use the Ball Chiu vertex as the longitudinal part of the
full vertex. In the next section, we discuss the elements which can serve as 
a guide in our hunt for the transverse piece of the full fermion-boson 
interaction.

\section*{Perturbation Theory}

       Only a correct
choice of the transverse vertex can lead to physically acceptable solutions.
How (and if) can one construct such a vertex? The only truncation of the 
complete set
of SDEs known so far that incorporates the key
features of a gauge theory such as the
WGT identities,
LKF transformations and gauge invariance of
physical observables, (e.g., the mass and the condensate)  at each level 
of approximation is perturbation theory. Therefore, it is natural to
assume that physically meaningful solutions
of the SDEs must agree with perturbative results
in the weak coupling regime. It requires, e.g., that every non-perturbative
{\em ansatz} chosen for the transverse vertex must reduce to its perturbative
counterpart when the interactions are weak. Perturbatively, the
transverse vertex is evaluated in the following fashion.
One evaluates the fermion propagator to a certain order and hence determines 
the longitudinal vertex to the same order. One also calculates perturbatively 
the full
vertex,  and a mere subtraction of the longitudinal part
yields the transverse part, the one which is not fixed by the WGTI.
A brief development of
work in this direction is outlined below~:
\begin{itemize}

\item Ball and Chiu, \cite{BC}, calculate one loop fermion boson vertex in 
Feynman gauge and hence propose a suitable basis to expand the transverse 
vertex.

\item Curtis and Pennington, \cite{CP}, calculate one loop fermion boson 
vertex in an
arbitrary covariant gauge in the limit when momentum in one of the fermion
legs is much greater than in the other. Using this as a guide, they
propose an {\em ansatz} for the transverse vertex involving just one basis 
vector and show 
that the gauge dependence of $\alpha_c$ is appreciably reduced.

\item Following the perturbative calculation in the first article of 
reference \cite{CP},
Bashir and Pennington, \cite{mythesis}, propose a vertex {\em ansatz} 
involving two 
basis vectors.
In terms of gauge invariance of the critical coupling, this {\em ansatz}
works much better than the Curtis-Pennington vertex and in a much 
wider range of values for the covariant gauge parameter.

\item  K{\i}z{\i}lers\"{u} {\em et. al.}, \cite{Ayse}, calculate complete one 
loop fermion boson vertex to ${\cal O}(\alpha)$ in an arbitrary covariant 
gauge and modify the
basis proposed by Ball and Chiu, \cite{BC}, to write out the transverse vertex.

\item Bashir {\em et. al}, \cite{BKP1}, calculate the perturbative constraint 
on the fermion boson vertex, imposed by the two
loop next to leading log calculation of the wavefunction renormalization.

\item Bashir and Raya, \cite{BR1}, calculate one loop fermion boson vertex 
in an arbitrary covariant gauge in 2+1 dimensions and, guided by it, propose
the first ever non-perturbative vertex which agrees with its 
full one loop expansion in the weak coupling regime.
This vertex has an explicit dependence 
on the gauge parameter $\xi$. They demonstrate, in the massless case, that a 
vertex cannot be constructed without an explicit dependence on $\xi$. 
For practical purposes of the numerical study of dynamical chiral
symmetry breaking, they also construct an effective 
vertex which shifts the angular dependence from the unknown fermion
propagator functions to the known basic functions, without changing its
perturbative properties at the one loop level. This
vertex should lead to a more realistic study of the dynamically generated
masses through the corresponding SDEs in 2+1 dimensions. 

\item Davydychev {\em et. al.}, \cite{DOS}, calculate the one loop vertex
in an arbitrary covariant gauge in arbitrary dimensions. This may help
one to construct a non perturbative vertex in arbitrary dimensions.

\end{itemize}

A two loop calculation of the transverse vertex would be useful, as it is 
likely to shed more light on its possible non-perturbative extensions.
We believe that a vertex which is reduced to its perturbative expansion
in the weak coupling regime stands a better chance to yield gauge 
invariant results.

\section*{Landau-Khalatnikov-Fradkin Transformations}

In a gauge field theory, Green functions transform in a specific 
manner under a variation of gauge, giving rise to LKF 
transformations in QED, \cite{LKF1}.
These were derived also by Johnson and Zumino through functional
methods, \cite{LKF2}. 
LKF transformations are non-perturbative in
nature and hence have the potential of playing an important role in 
addressing the problems of gauge invariance which plague the strong 
coupling studies of SDEs. In general, the 
rules governing these transformations are far from simple. The fact 
that they are written in coordinate space adds to their complexity. As 
a result, these transformations have played less significant and
practical role in the study of SDEs than desired.

 The LKF transformation for the three-point vertex is complicated and
hampers direct extraction of analytical restrictions on its
structure. Burden and Roberts, \cite{BCR1}, carried out a numerical 
analysis to compare the self-consistency of various {\em ansatze} for
the vertex, 
\cite{BC,CP,Haeri}, by means of its LKF transformation.
In addition to these numerical constraints, indirect analytical
insight can be obtained on the non-perturbative structure of the
vertex by demanding correct gauge covariance properties of the 
fermion propagator. References 
 \cite{CP,BKP1,BP,others} employ this idea. 
However, the inclusion of LKF transformations has been restricted to
massless fermions alone. The masslessness of the
fermions implies that the fermion propagator can be written only in
terms of one function, the so called wavefunction renormalization, $F(p)$.
In order to
apply the LKF transformation, one needs to know a Green function at least
in one particular gauge. This is a formidable task. However, one
can rely on approximations based on perturbation theory. It is
customary to take $F(p)=1$ in the Landau gauge, an
approximation justified by one loop calculation of the massless fermion
propagator in arbitrary dimensions, see for example, \cite{DOS}.
The LKF transformation then implies a power law for $F(p)$ in QED4
and a simple trigonometric function in QED3. To improve upon these
results, one can take two paths:
(i) incorporate the information contained in higher orders of
perturbation theory and (ii) study the massive theory. As pointed out
in \cite{BKP1}, in QED4, the power law structure of the wavefunction
renormalization remains intact by increasing order of
approximation in perturbation theory although the exponent of course
gets contribution from next to leading logarithms and so 
on\footnote{For the two loop calculation of the fermion propagator,
see for example \cite{RF1}.}. In \cite{BKP1}, constraint was obtained
on the 3-point vertex by considering a power law where the exponent 
of this power law was not restricted only to the one loop fermion 
propagator. In QED3, the two loop fermion propagator was evaluated in
\cite{B2loop}, where it was explicitly shown that the 
the approximation $F(p)=1$ is only valid upto one loop, thus violating
the {\em transversality condition} advocated in the second article of
reference \cite{others}. The result
found there was used used in \cite{adnan} to find the improved
LKF transform. Later on, in reference \cite {BR2}, the LKF transformed 
fermion propagator in massive QED3 and QED4 was evaluated
with the  simplest input which
corresponds to the lowest order of perturbation theory, i.e.,
the propagator being bare in the Landau gauge. We believe that the 
incorporation of LKF transformations, along with WGT identities, in the SDE
can play a key role in addressing the problems of gauge invariance. For
example, only those assumptions should be permissible which keep intact the
correct behavior of the Green functions under the LKF transformations, in
addition to ensuring that the WGTI is satisfied. It makes it vital to explore
how two and three-point Green functions transform in a gauge covariant
fashion. 

\section*{Other Key Factors}

We now comment on other key factors which should be taken into account
while studying SDEs~:

\begin{itemize}

\item

The transverse vertex should be free of any kinematic singularities.
Within the framework of the basis proposed by K{\i}z{\i}lers\"{u} 
{\em et. al.}, \cite{Ayse}, it amounts to saying that every coefficient of
the basis vectors itself should be free of kinematic singularities.

\item

As discovered by Curtis and Pennington, \cite{CP}, multiplicative 
renormalizability
of the fermion propagator plays an important role in the restoration of
 gauge invariance of the critical coupling above which masses are 
generated for fundamental fermions. However, their work as well as the one
presented in reference \cite{BP}, only incorporates the leading
log behavior of the propagator in the construction of the non-perturbative
vertex.  However, in a
subsequent work, Bashir {\em et. al.} presented the most general 
construction of the transverse vertex required by 
multiplicative renormalizability of the fermion propagator to all
orders, \cite{BKP1}.

\item

 If one takes into account all the relevant features 
mentioned so far, one is likely to acquire gauge invariance of all the
physical observables. However, it is a prohibitively difficult
to implement all the constraints to the required degree. Therefore, a
direct requirement of the gauge invariance of the physical observables
can serve as an additional driving force to constrain the fermion-boson
vertex. One such attempt is made by  Bashir {\em et. al.}, \cite{BP}.
They hold the critical coupling to be gauge invariant and obtain 
constraints on the transverse vertex.

\item

The works described so far use cut-off regularization scheme 
to study the gauge dependence of the physical observables related to
the fermion propagator. As the cut-off method in general does not respect 
gauge symmetry, a criticism of these works has been raised recently, 
\cite{DRS}. They suggest dimensional regularization scheme to study the 
chirally asymmetric phase of QED so that the possible gauge dependence 
coming from the inappropriate regulator could be filtered out.
However, implementation of dimensional regularization leads to complicated
kernels in the coupled integral equations which are then hard to
solve, \cite{DRS,BHR}.

\end{itemize}

\section*{Conclusions}

       We summarize the attempts made so far to make the study of 
Schwinger-Dyson equation for the fermion propagator in QED more
realistic by constructing an {\em ansatz} for the fermion boson interaction
in such a fashion that it can effectively recuperate the necessary
information lost on truncating the infinite tower of these equations.
Although a lot of work has been carried out in this direction, more work is
needed to make these studies fully reliable. One should then embark on 
the studies of unquenched QED and move on to consider more realistic
cases such as QCD and the improved versions of top quark condensation.

\section*{Acknowledgments}

We acknowledge the CIC and the CONACyT grants under the projects 4.12 and 
32395-E, respectively.

\end{document}